\documentclass[a4paper,11pt]{article}
\usepackage{jheppub} 
\usepackage{lineno}

\usepackage{graphicx}
\usepackage{dcolumn}
\usepackage{bm}

\usepackage[T1]{fontenc}  
\usepackage[normalem]{ulem} 
\usepackage{graphicx}
\usepackage{tikz}       
\usepackage{subcaption}
\usepackage{diagbox}

\def\be{\begin{equation}}
\def\ee{\end{equation}}
\def\ba{\begin{eqnarray}}
\def\ea{\end{eqnarray}}

  \def\t {\tau}     \def\a {\alpha}  \def\d {\dot}  \def\g {\gamma}        \def\b {\beta}      
              \def\.{\cdot}

\def\mR{\mathcal{R}}

\title{Thermodynamics of the FRW Universe in Generalized Proca Theory}

\author[1]{Aofei Sang\note{Corresponding author.}}
\affiliation{${}^1$Department of Physics, Southern University of Science and Technology, Shenzhen 518055, China}
\emailAdd{aofeisang@mail.bnu.edu.cn}
\emailAdd{12331027@mail.sustech.edu.cn}

\abstract{We investigate the thermodynamics of a spatially flat Friedmann-Robertson-Walker (FRW) universe within the framework of Generalized Proca (GP) theory, a comprehensive vector-tensor theory. By adopting two distinct dark energy models in GP, we derive the corresponding modified Friedmann equations, the thermodynamic first law for the apparent horizon, and the equation of state for these models. For the first model, characterized by power-law couplings and an ansatz linking the Proca field to the Hubble parameter, we analytically demonstrate the existence of a critical point in the pressure-volume$(P-V)$ diagram, indicating a $P-V$ phase transition. For the second model, defined by a Proca field marginal coupled to curvature, we show that there is no phase transition when the coupling constants are selected within the range permitted by observations. This investigation not only extends the FRW thermodynamics to vector-tensor theories but also demonstrates that cosmological phase transitions can serve as a powerful diagnostic tool for distinguishing viable dark energy models in GP theory.

}

\begin{document}
\maketitle
\flushbottom

\section{Introduction}\label{sec1}




The current standard cosmological model, the Lambda-Cold-Dark-Matter($\Lambda$CDM) model, has achieved remarkable success in explaining a wide range of observational data. Built upon Einstein’s general relativity, it assumes that the universe is dominated by CDM and a cosmological constant $\Lambda$ with constant energy density. This model provides an excellent fit to key observational results, including the anisotropies of the cosmic microwave background (CMB), the formation of large-scale structure, the luminosity distance–redshift relation of Type Ia supernovae, and baryon acoustic oscillations (BAO)\cite{SupernovaCosmologyProject:1998vns,SupernovaSearchTeam:1998fmf,WMAP:2003elm,Planck:2013pxb}. Nevertheless, the $\Lambda$CDM model faces serious theoretical challenges. The most prominent among these is the cosmological constant problem: the value of $\Lambda$ inferred from observations is smaller than the vacuum energy predicted by quantum field theory by approximately $120$ orders of magnitude\cite{Weinberg:1988cp}. This discrepancy has motivated the exploration beyond $\Lambda$CDM.

One prominent approach is to introduce a dynamical mechanism to replace the static cosmological constant. Among various proposals, scalar–tensor theories have attracted considerable attention due to their simplicity and rich dynamical behavior\cite{Horndeski:1974wa,Clifton:2011jh}. In particular, the covariant Galileon theory\cite{Deffayet:2009wt,Deffayet:2009mn,Deffayet:2011gz,Horndeski:1974wa}—a subclass of scalar–tensor theories that yields second-order equations of motion and represents one of the most general ghost-free frameworks—successfully produces self-accelerating cosmological solutions by incorporating higher-derivative self-interaction terms that remain free of ghost instabilities\cite{Silva:2009km,DeFelice:2010pv,Kobayashi:2010cm,Burrage:2010cu}. This theory not only provides a viable candidate for dynamical dark energy but also exhibits the Vainshtein screening mechanism in certain parameter regimes, thereby recovering the predictions of general relativity on solar system scales\cite{Vainshtein:1972sx}.
A natural question then arises: can a similar construction be extended to vector fields? However, for massless vector fields that preserve U(1) gauge symmetry, a powerful no-go theorem forbids the introduction of nontrivial higher-derivative self-interactions analogous to those in Galileon theories without introducing ghost instabilities\cite{BeltranJimenez:2013btb,Deffayet:2016von}. This implies that, in order to construct a modified gravity theory based on vector fields, one must abandon $U(1)$ gauge symmetry by introducing a mass term for the vector field, leading to the Proca field, a massive spin-1 field possessing three physical degrees of freedom.
In ref.\cite{Heisenberg:2014rta}, the Generalized Proca (GP) theory was systematically formulated. This framework extends the standard Proca action by adding a series of carefully designed higher-derivative self-interaction terms. These terms are constructed such that the resulting equations of motion remain at most second order, while guaranteeing that only three healthy physical degrees of freedom. In cosmology, GP theory demonstrates remarkable explanatory power: its time component $A_0(t)$ can act as a dynamical source of dark energy, driving the late-time accelerated expansion of the universe. Moreover, GP models can satisfy current observational constraints from the CMB, SNIa, and BAO, while potentially leaving distinctive imprints on primordial gravitational waves that could be detectable in future observations.\cite{DeFelice:2016uil,DeFelice:2016yws,Savaliya:2025cct,Garcia-Serna:2025hpt} Therefore, Generalized Proca theory not only offers a novel theoretical avenue for addressing the mystery of dark energy but also provides a testable alternative framework for upcoming high-precision cosmological surveys.

Parallel to the development of modified gravity, a deep connection between gravitation and thermodynamics has been uncovered over the past few decades. In the 1970s, Bekenstein and Stephen Hawking systematically formulated the four laws of black hole mechanics, revealing that black holes are physical systems endowed with a rich thermodynamic structure\cite{Hawking:1974rv,Hawking:1975vcx,Bekenstein:1973ur,Bekenstein:1974ax,Bardeen:1973gs}. Central to this framework is the first law, which relates changes in a black hole’s mass to variations in its horizon area, angular momentum, and charge, and establishes that the black hole temperature is proportional to its surface gravity while its entropy is proportional to the horizon area. Since the proposal of black hole mechanics, it has rapidly become one of the core fields in theoretical physics research. Extensive work has focused on static or stationary black holes, systematically constructing thermodynamic descriptions across various black hole solutions and gravitational theories\cite{Wald:1993nt,Yang:2025rud,Sadeghi:2025por,Kubiznak:2012wp,Ahmed:2025dst,Jiang:2021pna,Jiang:2025aqw,Sang:2021rla,Sang:2024tjd}.
Black hole thermodynamics has thus provided crucial insights into the nature of gravity and the microscopic origin of spacetime.
However, the universe is far from stationary—whether it be black holes accreting matter, undergoing mergers, or the cosmic expansion itself, all represent examples of dynamical spacetime backgrounds. In such settings, basic concepts of traditional black hole thermodynamics—such as the event horizon and stationary surface gravity—face serious challenges. First, the event horizon is globally defined, relying on the full causal structure of spacetime, making it impractical for local physical processes or observational purposes. Second, in dynamical spacetimes, surface gravity is generally not constant across the horizon, making the notion of temperature ambiguous. 
To address these difficulties, researchers have shifted toward using local or quasi-local geometric constructs instead of the global event horizon. Among the most influential approaches are Hayward’s dynamical horizon\cite{Hayward:1994bu,Hayward:1993wb,Hayward:1997jp,Hayward:1998ee}. In cosmology, the apparent horizon—defined as the surface where the expansion of outgoing null geodesics vanishes—has become the natural candidate for a thermodynamic horizon\cite{Cai:2005ra,Cai:2006pa,Cai:2006rs,Cai:2008gw,Cai:2008mh,Abdusattar:2021wfv,Abdusattar:2023hlj}. In ref.\cite{Jacobson:1995ab}, Jacobson further demonstrated that if one assumes the Clausius relation holds for arbitrary local Rindler horizons (accelerated causal boundaries), Einstein’s field equations can be derived as an equation of state. This not only provided compelling evidence for a thermodynamic origin of gravity but also charted a clear path toward formulating a consistent thermodynamic description in dynamical settings.
Since then, this paradigm has been successfully extended to numerous modified theories of gravity. By redefining entropy and energy, the first law of thermodynamics has been reconstructed in dynamic cosmological contexts\cite{Abdusattar:2021wfv,Abdusattar:2022bpg,Abdusattar:2023hlj,Chu:2025zuz,Kong:2022xny}. This thermodynamic perspective has thus opened new avenues for understanding cosmic acceleration, the nature of dark energy, and even inflation in the early universe.

An intriguing question is the thermodynamics of FRW universe in the context of GP theory. While black hole thermodynamics in GP theory has received preliminary investigation\cite{Minamitsuji:2024ygi}, its thermodynamic behavior in the more compelling and widely studied cosmological setting remains unexplored. In this paper, we will systematically study the thermodynamics of an FRW universe within GP theory. We will assign entropy and temperature to the cosmological apparent horizon and examine whether the first law of thermodynamics holds. We further probe for possible interesting phase transitions and critical phenomena.
Ultimately, cosmological phase transitions may hold clues to cosmic history.

The paper is structured as follows: In Section \ref{sec2}, we outline the dynamics of the FRW universe in GP theory, presenting the relevant action and field equations. In Section \ref{sec3}, we give a brief review of the FRW thermodynamics in Einstein gravity. In Section \ref{sec4}, we delve into the thermodynamics for two chosen models. 
Finally, we conclude in Section \ref{sec5} with a summary of our findings.
We use $(-,+,+,+)$ and the Planck unit with $c=\hbar=k_B=1$ in this paper.


\section{Dynamics of the FRW Universe in Generalized Proca Theory}\label{sec2}

In this paper, we consider the Generalized Proca theory with the action \cite{DeFelice:2016uil,DeFelice:2016yws}
\be\begin{aligned}\label{action}
S=\int d^4x \sqrt{-g} \left(\mathcal{L}_m+ \sum_{i=2}^5 \mathcal{L_{i}} \right)\,.
\end{aligned}\ee
The Lagrangian with the non-minimally coupled parts reads
\be\begin{aligned}
\mathcal{L}_2=&G_2(X,\mathcal{F},\mathcal{G})\,,\\
\mathcal{L}_3=&G_3(X)\nabla_a A^a\,,\\
\mathcal{L}_4=&G_4(X)R+G_{4,X}(X)\left[(\nabla A)^2+c_1 \nabla_aA_b \nabla^a A^b-(1+c_1) \nabla_aA_b \nabla^b A^a\right]\,,\\
\mathcal{L}_5=&G_5(X)G_{ab}\nabla^a A^b-\frac{1}{6}G_{5,X}(X)\left[(\nabla A)^3-3 c_2 (\nabla A) \nabla_aA_b \nabla^a A^b -3(1-c_2)(\nabla A)\nabla_aA_b \nabla^b A^a\right.\\
&\left.+(2-3c_2)\nabla_a A_b \nabla^c A^a \nabla^b A_c+3 c_2 \nabla_a A_b \nabla^c A^a \nabla_c A^b \right]\\
\mathcal{L}_6=&G_6(X)\tilde{R}^{abcd}\nabla_a A_b \nabla_c A_d+\frac{1}{2}G_{6,X}(X)\tilde{F}^{ab}\tilde{F}^{cd} \nabla_a A_c \nabla_b A_d
\end{aligned}\ee
with the following definition
\be\begin{aligned}
    &X=-\frac{1}{2}A^a A_a\,,\,\,\, \mathcal{F}=-\frac{1}{4}F_{ab}F^{ab}\,,\,\,\, \mathcal{G}=A^a A^b F_a{}^c F_{bc}\,,\\
    &\nabla A=\nabla_a A^a\,,\,\,\,\tilde{R}^{abcd}=\frac{1}{4}\epsilon^{aba_1 b_1}\epsilon^{cdc_1 d_1}R_{a_1b_1c_1d_1}\,,\,\,\, \tilde{F}^{cd}=\frac{1}{2}\epsilon^{abcd} F_{ab}\,.
\end{aligned}\ee
$F_{ab}=\nabla_a A_b-\nabla_b A_a$ is the strength tensor of the Proca field. $G_i$ are arbitrary functions and $G_{i,X}(X)\equiv\partial G_i(X)/\partial X$. $c_1$ and $c_2$ are arbitrary constants.  The Lagrangian we showed here is the most general vector-tensor theory, whose equations of motion contain only second-order derivatives, thereby naturally avoiding Ostrogradsky ghost instabilities. In fact, we can find later that the value of $c_1,c_2$ and the form of $G_6(X)$ will not have any effect on the equation of motion in the FRW universe. When $G_2=-m^2 X+\mathcal{F}$ and $G_4=M_{pl}^2/2$, one can recover Einstein gravity minimally coupled with Proca fields, where $m$ is the mass of the Proca field and will lead to $U(1)$ symmetry breaking if not vanish and $M_{pl}^2\equiv 1/ (8\pi G)$ is the square of Planck mass with dimension $[L^{-2}]$.

Another term in \eqref{action}, $\mathcal{L}_m$, represents the Lagrangian of the minimally coupled matter field, where we assume it to be a perfect fluid in this paper. The energy momentum of the matter field is defined by
\be\begin{aligned}
    T_{ab}=-\frac{2}{\sqrt{-g}}\frac{\sqrt{-g} \delta\mathcal{L}_m}{\delta g^{ab}}\,.
\end{aligned}\ee
For the perfect fluid with energy density $\rho_m$ and pressure $p_m$, its stress-energy tensor can be written as 
\be\begin{aligned}\label{perfectfluidT}
    T_{ab}=(\rho_m+p_m)U_a U_b+p_m g_{ab}\,,
\end{aligned}\ee
where $U^a$ is the $4$-velocity of the perfect fluid. The stress-energy tensor should satisfy the continuity equation 
\be\begin{aligned}
    \nabla_a T^{ab}=0\,.
\end{aligned}\ee

Considering the isotropic and homogeneous nature of the universe, the metric and the Proca field need to satisfy the ansatz:
\be\begin{aligned}\label{metricandproca}
    &ds^2=-dt^2+a(t)^2 dr^2+\mR(t,r)^2d\Omega^2\\
    &A_a=A_t(t)(dt)_a
\end{aligned}\ee
In this paper, we will focus on the flat universe with $k=0$, and we define $\mathcal{R}(t,r)=r a(t)$.
Besides, in order to be consistent with Ref.\cite{DeFelice:2016yws,BeltranJimenez:2016rff}, we use a scalar function $\phi(t)\equiv A^{t}(t)=-A_0(t)$ to describe the evolution of the Proca field. 
For the minimally coupled matter field, using $U^a=(\partial/\partial t)^a$, we can obtain $T^{\mu\nu}=\text{diag}(\rho_m, \, p_m/a^2,\, p_m/\mathcal{R}(t,r)^2,\, \csc^2\theta\, p_m/ \mathcal{R}(t,r)^2 )$
where the Greek letter represent for the component in $\{t,r,\theta,\varphi \}$ coordinate system. Then, we can write the continuity equation explicitly as
\be\begin{aligned}\label{continuity}
    \dot{\rho}_m+3 H (\rho_m+p_m)=0\,,
\end{aligned}\ee
where $H=\dot{a}/a$ is the Hubble parameter and $\dot{}$ is for derivative with respect to $t$.

Varying the Lagrangian with respect to $g_{tt}$, $g_{rr}$, and $A_t$, we can get the Friedmann equation and equation of motion for $\phi$:
\be\begin{aligned}\label{friedmann}
\rho_m=& G_2-G_{2,X}\phi^2-3 G_{3,X} H\phi^3+6H^2\left(G_4-2G_{4,X}\phi^2-G_{4,XX}\phi^4\right)+H^3\phi^3\left(5G_{5,X}+G_{5,XX}\phi^2\right)\,,\\
p_m=&-G_2+G_{3,X}\phi\dot{\phi} -2G_4(3H^2+2\dot{H})+2G_{4,X}\phi (3H^2 \phi+2\d{H}\phi+2H\d{\phi})+4G_{4,XX}H\phi^3\d{\phi}\\
&-G_{5,X}H\phi(2H^2\phi+2\d{H}\phi+3H\d{\phi})-G_{5,XX}H^2\phi^4\d{\phi}\,,
\end{aligned}\ee
and
\be\begin{aligned}\label{eomA}
0=&\phi \left[G_2+3G_{3,X}H \phi+6H^2(G_{4,X}+\phi^2 G_{4,X})-H^3(3G_{5,X}\phi+G_{5,XX}\phi^3) \right]\,,    
\end{aligned}\ee
which are also derived in Ref.  
Eq. \eqref{eomA} is an algebraic equation of $\phi$. Therefore, we can get $\phi$ without solving the differential equation. This implies that there is no extra degree of freedom propagated by the Proca field. We would like to focus on the nontrivial branch with $\phi\neq 0$ in this paper. Also note that there are only three independent equations between \eqref{continuity}, \eqref{friedmann}, and \eqref{eomA}.

\section{Review of FRW Thermodynamics}\label{sec3}

In this section, we would like to give a brief review of the FRW thermodynamics in Einstein gravity minimally coupled with the perfect fluid. The basic ansatz is the same as that in sec.\ref{sec2}, where the metric is \eqref{metricandproca} and the stress-energy tensor is \eqref{perfectfluidT}.


In dynamic spacetime, thermodynamics is formulated on the apparent horizon rather than the event horizon because the former is locally defined, always exists, and respects causality. It depends only on instantaneous geometric quantities, and yields a thermodynamic first law consistent with the Friedmann equations\cite{Hayward:1993wb,Hayward:1994bu,Hayward:1997jp,Hayward:1998ee,Cai:2006rs}. 
In FRW spacetime, the metric \eqref{metricandproca} can be decomposed as
\be\begin{aligned}
    g_{ab}=-\xi_a k_b-k_a\xi_b+h_{ab}\,,
\end{aligned}\ee
where $h_{ab}$ is the induced metric on the $2$-sphere, 
\be\begin{aligned}
    \xi_a=\frac{1}{\sqrt{2}}\left[(dt)_a+a(t)(dr)_a\right]
\end{aligned}\ee
is the outgoing null vector, and 
\be\begin{aligned}
    n_a=\frac{1}{\sqrt{2}}\left[(dt)_a-a(t)(dr)_a\right]
\end{aligned}\ee
is the ingoing null vector.
The apparent horizon $\mR_A$ is defined by the marginally trapped horizon with vanishing expansion, which means $\mR_A$ is determined by $h_{ab}\nabla^a\xi^b=0$. Then, it is easy to get
\be\begin{aligned}\label{RtoH}
 \mR_A=\frac{1}{H}\,,\,\,\,\,\,.   
\end{aligned}
\ee
The time derivative of \eqref{RtoH} gives
\be\begin{aligned}\label{RdtoHd}
\d \mR_A=-\frac{\d H}{H^2}    \,,
\end{aligned}
\ee
which is useful in the following discussion.
Then, we can defined the temperature of the apparent horizon\cite{Abdusattar:2021wfv,Abdusattar:2022bpg,Abdusattar:2023hlj,Cai:2008gw,Cai:2008mh}
\be\begin{aligned}\label{temp}
  T\equiv\frac{|\kappa|}{2\pi}=\frac{1}{2\pi \mR_A}\left(1-\frac{\d \mR_A}{2}\right)  
\end{aligned}
\ee

In Einstein's gravity, we also get the continuity equation
\be\begin{aligned}\label{sec3-conti}
    \d\rho_m +3 H(\rho_m+p_m)=0\,.
\end{aligned}\ee
Define the total energy\cite{Cai:2006rs,Abdusattar:2023hlj}
\be\begin{aligned}\label{totalenergy}
  E=\rho_m V \,, 
\end{aligned}
\ee
where $V\equiv\frac{4\pi}{3}\mR_A^3$. 
Then, the total differential of $E$ gives
\be\begin{aligned}\label{dE}
  dE &=V d\rho_m+\rho_m dV\\
  &=-4\pi \mR_A^2(\rho_m+p_m)dt+\rho_m dV\\
  &=-4\pi \mR_A^2(\rho_m+p_m)\left(1-\frac{\d \mR_A}{2}\right) dt+W dV\\
  &=-8\pi^2 \mR_A^3(\rho_m+p_m) T dt+W dV\,,
\end{aligned}
\ee
where we used the continuity equation \eqref{sec3-conti} and the relation between $H$ and $\mR_A$ \eqref{RtoH} for the second equal sign. Besides, we used $dV=4\pi\mR_A^2\d \mR_A dt$ for the third equal sign and the expression of temperature \eqref{temp} for the last equal sign. And, the work density $W$ is defined by
\be\begin{aligned}\label{workdensity}
 W\equiv \frac{1}{2}(\rho_m-p_m)\,,   
\end{aligned}
\ee
So far, we have not used the specific Friedmann equation, and the derivation is valid for any gravitational theories.

Then, using the Friedmann equations
\be\begin{aligned}\label{sec3-freid}
    &\rho_m=3 M_{pl}^2 H^2\,,\\
    &p_m=-M_{pl}^2\left(2\d H +3H^2\right)\,,
\end{aligned}\ee
and \eqref{sec3-conti}, we can find 
\be\begin{aligned}
  \rho_m+p_m=-\frac{\d\rho_m}{3H}=  -2M_{pl}^2 \d H\,.
\end{aligned}
\ee
Here, $M_{pl}^2=1/(8\pi G)$ is the square of Planck mass with dimension $[L^{-2}]$.
Substituting this into the first term of \eqref{dE} and replacing $\d H$ using \eqref{RdtoHd}, we can get
\be\begin{aligned}\label{firstlaw0}
    dE&=T \left(2\pi\mR_A^3 \d H\right)dt+W dV\\
    &=-T dS+WdV\,.
\end{aligned}\ee
with $S=\pi \mR_A^2$. 
The expression for $S$ is picked up from the identity $dE=-TdS+WdV$, which agrees with the expression in black hole thermodynamics.
For modified gravity, $S$ usually needs to be modified. Eq. \eqref{firstlaw0} is the thermodynamic first law of the FRW universe in Einstein gravity.

\section{Thermodynamics of the FRW Universe in Generalized Proca Theory}\label{sec4}

With the previous preparations, we can investigate the thermodynamics of FRW spacetime under the framework of GP theory in this section.
In this paper, we would like to focus on two interesting dark energy models.  
One is discussed in ref.\cite{DeFelice:2016uil,DeFelice:2016yws}, and the other is discussed in ref.\cite{Savaliya:2025cct}.
For both models, the Proca fields can serve as dark energy to accelerate the expansion of the universe.

\subsection{Model A}\label{sec4-1}

Model A assumes a power-law relation between the time component $\phi$ of the Proca field and the Hubble parameter $H$, and accordingly constructs a GP theory that realizes this relation. With appropriately chosen parameters, the resulting cosmological model naturally undergoes radiation- and matter-dominated epochs and asymptotically approaches a stable de Sitter solution, thereby successfully accounting for the current accelerated expansion of the universe. At the perturbative level, the model remains stable throughout the entire cosmic history within a reasonable range of parameters.

Following \cite{DeFelice:2016uil,DeFelice:2016yws}, we would like to assume 
\be\begin{aligned}\label{model1phi}
    \phi^p=\lambda^p H^{-1}
\end{aligned}\ee
where $\lambda$ is a constant with dimension $\lambda^p\sim [L]^{-p-1}$.
To ensure that $\phi$ and $H$ evolve over time and not vanish, we have to require $\lambda\neq 0$ and $p\neq 0$. We can go back to the Einstein FRW universe by letting $\lambda=0$, which means the Proca field vanishes.
This can be achieved by choosing the power-law coupling functions.
We would like to choose
\be\begin{aligned}\label{model1G}
        G_2=\mathcal{F}+b_2 X^{p_2}\,,\,\,G_3=b_3 X^{p_3}\,, \,\,G_4=\frac{M_{pl}^2}{2}+b_4 X^{p_4}\,, \,\,G_5=b_5 X^{p_5}\,.
\end{aligned}\ee
It is obvious that $p_{2,3,4,5}$ are dimensionless while $b_{2,3,4,5}$ are dimensional. 
Pluging \eqref{model1phi} and \eqref{model1G} into \eqref{eomA} and considering the equation should be time independent, the power of $H(t)$ need to be $0$. Then, we can find $3$ relations between $p$ and $p_{2,3,4,5}$, which allow us to express $p_{3,4,5}$ using $p$ and $p_2$ as
\be\begin{aligned}
    p_3=\frac{1}{2}(p+2p_2-1)\,,\,\,\,p_4=p+p_2\,,\,\,\, p_5=\frac{1}{2}(3 p+2 p2-1)\,.
\end{aligned}\ee
It is convenient to define dimensionless quantities
\be\begin{aligned}
    \beta_2=\frac{ b_2}{3\cdot 2^{p_2}M_{pl}^2}\lambda^{2p_2}H_0^{-2-\frac{2p_2}{p}}\,,\,\,\beta_i=\frac{p_i b_i(\lambda^p)^{i-2}}{2^{p_i-p_2}p_2 b_2}\,,\,\,s=\frac{p_2}{p}\,
\end{aligned}\ee
with $i=3,4,5$. Here, $H_0$ is the Hubble parameter today.  
Observational constraints from SNIa, CMB, and BAO data limit the parameter to $0\leq s \leq 0.36$(see Ref.\cite{DeFelice:2016yws}).
Additionally, requiring that the dark energy density parameter $\Omega_{\text{DE}}$ remains finite imposes the condition $1+s>0$\cite{DeFelice:2016yws}.
Using these definitions, \eqref{eomA} reduce to
\be\begin{aligned}
    1+3\beta_3+6(-1+2p+2p_2)\beta_4-3 p\beta_5-2p_2\beta_5=0\,,
\end{aligned}\ee
which gives $\beta_3=\frac{1}{3}(-1+6\beta_4-12 p \beta_4-12 p_2\beta_4+3p \beta_5+2p_2\beta_5)$. With the expression of $\beta_3$, \eqref{friedmann} gives
\be\begin{aligned}\label{rhopA}
    &\rho_m=3M_{pl}^2 H^2+\frac{C_1 M_{pl}^2}{(p+p_2)} H^{-2s}\,,\\
    &p_m=-M_{pl}^2(3H^2+2\d{H})+\frac{C_2 M_{pl}^2 H^{-2s}}{ (p+p_2)}\left(3-2 s \frac{\d{H}}{H^2}\right)\,.\\
\end{aligned}\ee
$C_{1,2}$ are constants composed by $p$, $p_2$, $\beta_{2,4,5}$, and $H_0$, which is fixed once the theory is fixed. Moreover, the current astronomical observations impose no restrictions on $C_1$. The expressions of $C_i$ are shown in the Appendix \ref{app}. Since $M_{pl}^2$ is an overall factor, it is safe and convenient to set $G=1$ in this subsection. In the following discussion in sec.\ref{sec4-1}, we would like to use $G=1$.
It can be easily checked that $C_{1,2}=0$ and the Friedmann equation reduces to Einstein gravity when $\beta_{2,4,5}=0$.

Then, we can calculate the thermodynamic quantities given in sec.\ref{sec2}.
As discussed in sec.\ref{sec2}, we work on the apparent horizon \eqref{RtoH}. 
According to the definition of the total energy eq.\eqref{totalenergy} and work density eq.\eqref{workdensity}, combining with \eqref{rhopA}, replacing $H$ with $\mathcal{R}_A$ using \eqref{RtoH} and \eqref{RdtoHd}, we can obtain the equality on the apparent horizon 
\be\begin{aligned}\label{dEmodel1}
    dE=-T dS+W dV
\end{aligned}\ee
with
\be\begin{aligned}\label{thermo1}
    &E=\frac{\mathcal{R}_A}{2}+C_1 \frac{1}{6(p+p_2)}\mR_A^{3+2s} \,,\\
    &S=\pi \mR_A^2-C_1\frac{\pi p_2}{3(p+p_2)(2p+p_2)}\mR_A^{4+2s}\,,\\
    &W=\frac{3-\d{\mR}_A}{8\pi \mR_A^2}+C_1 \frac{(3+s \d{\mR}_A)}{24\pi (p+p_2)} \mR_A^{2s}\,,\\
    &T=\frac{1}{2\pi \mR_A}\left(1-\frac{\d{\mR}_A}{2}\right)\,,\,\,\,V=\frac{4}{3}\pi \mR_A^3\,.
\end{aligned}\ee
Terms proportional to $C_1$ contain all corrections given by the nonminimally coupled Proca field. If we let $C_1=0$, which can be achieved when $\lambda=0$, i.e., the Proca field vanishes, the results \eqref{thermo1} are exactly the same as thermodynamic quantities in Einstein gravity.

From \eqref{dEmodel1} and the first law $dU=T dS-PdV$, we can identify the internal energy $U=-E$ and pressure $P=W$.
Then, we can find the equation of state by expressing $P$ with $T$ and $V$(or $\mR_A$)
\be\begin{aligned}\label{eos1}
    P(T,\mR_A)=\frac{4 \pi  \mR_A T+1}{8 \pi  \mR_A^2}+C_1 \frac{ \left(-4 \pi  p_2 \mR_A T+3 p+2 p_2\right)}{24 \pi  p \left(p+p_2\right)} \mR_A^{2s}\,.
\end{aligned}\ee
When $s=-3$, the equation of state is the same as that of the effective scalar-tensor theory, which is studied in ref.\cite{Abdusattar:2023hlj}. 
Next, we would like to focus on a more physical case where $p=5/2$ and $p_2=1/2$ and study the phase transition and critical phenomenon.

We can determine whether there exist the $P-V$ phase transition of the FRW universe in GP theory by calculating the critical point, which is defined by the inflection point of $P(V)$.
Combining \eqref{eos1} and 
\be\begin{aligned}
    0=\left(\frac{\partial P}{\partial \mR_A}\right)_T=    \left(\frac{\partial^2 P}{\partial \mR_A^2}\right)_T\,,
\end{aligned}\ee
we can acquire the critical point
\be\begin{aligned}\label{criticalpoint}
    T_c=\alpha_1 C_1^{\frac{5}{12}}\approx 0.037273 C_1^{\frac{5}{12}}\,,\,\,\, \mR_c=\alpha_2 C_1^{-\frac{5}{12}}\approx 6.6543 C_1^{-\frac{5}{12}}\,,\,\,\,P_c=\alpha_3 C_1^{\frac{5}{6}}\approx 0.029898 C_1^{\frac {5}{6}}\,
\end{aligned}\ee
where the exact value of $\a_1$, $\a_2$ and $\a_3$ are shown in the Appendix \ref{app}.
Since $T$ and $\mR_A$ should be positive real numbers, the critical point only exists when $C_1>0$.
To investigate the behavior near the phase transition point clearly, we define the dimensionless thermodynamic quantities by
\be\begin{aligned}\label{dimensionlessPTVR}
    \t=\frac{T}{T_c}\,,\,\,\, r_A=\frac{\mR_A}{\mR_c}\,,\,\,\, p=\frac{P}{P_c}\,,\,\,\, v=\frac{V}{V_c}=r_A^3\,.
\end{aligned}\ee
Then, the equation of state \eqref{eos1} gives
\be\begin{aligned}\label{dimensionlesseos}
    p=\frac{1}{8 \pi  \alpha _2^2 \alpha _3 r_A^2}+\frac{\alpha _1 \tau }{2 \alpha _2 \alpha _3 r_A}+\frac{17 \alpha _2^{2/5} r_A^{2/5}}{360 \pi  \alpha _3}-\frac{\alpha _1 \alpha _2^{7/5} \tau  r_A^{7/5}}{90 \alpha _3}\,,
\end{aligned}\ee
which is independent of the selection of the coupling constant of the theory.

\begin{figure}
\centering 
\includegraphics[width=0.7\textwidth]{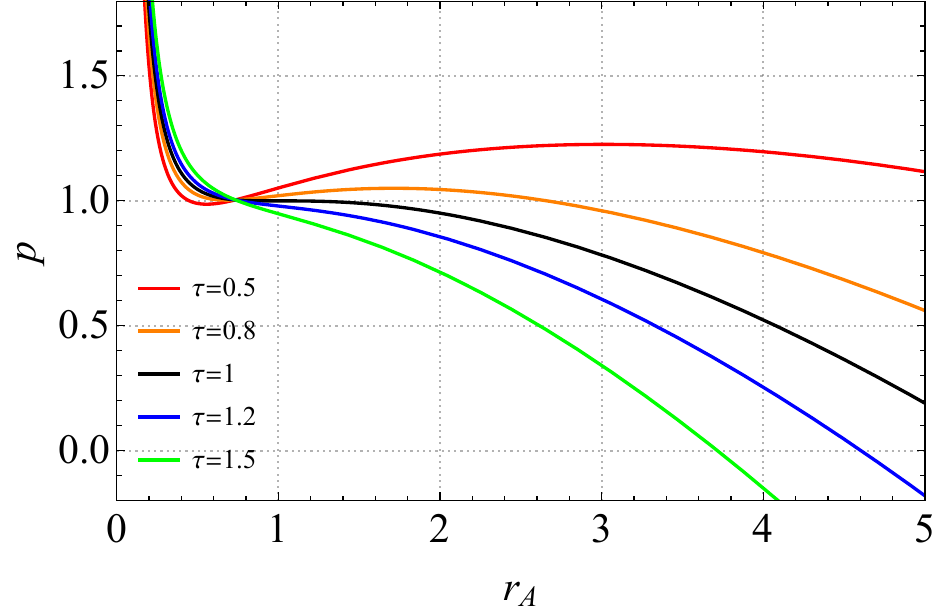}
\caption{Isothermal curves of the dimensionless pressure $p$ versus dimensionless AH $r_A$ for Model A in GP cosmology. Different colors correspond to different dimensionless temperatures: $\t=0.5$(Red), $\t=0.8$(Orange), $\t=1$(Black, critical isotherm), $\t=1.2$(Blue), $\t=1.5$(Green). The characteristic van der Waals–like loop appears for $\t<1$, while the critical isotherm shows a clear inflection point at (1,1).}
\label{pv-1} 
\end{figure}

Fig.\ref{pv-1} displays the isothermal curves of the dimensionless pressure $p$ as a function of the apparent horizon radius for the FRW universe in Model A of GP theory. Each curve corresponds to a distinct dimensionless temperature $\t$. For temperatures above the critical value$(\t>1)$, the isotherms are monotonically decreasing, reflecting a stable, gas-like phase. At the critical temperature$(\t=1)$, the isotherm exhibits an inflection point. Below the critical temperature $\t<1$, the isotherms develop a “van der Waals loop”, indicating thermodynamic instability in the intermediate region and the coexistence of two distinct phases(analogous to liquid and gas). This behavior strongly suggests the presence of a first-order phase transition in the cosmological evolution governed by this model. Moreover, the appearance of negative dimensionless pressure in the isothermal curves reflects the effective dark energy behavior or repulsive gravitational effects inherent in GP cosmology, consistent with the observed accelerated expansion of the universe.

Next, we would like to study the critical exponents in this model. The $4$ critical exponents, $\tilde{\a}$, $\tilde{\b}$, $\tilde{\g}$ and $\tilde{\delta}$ , are defined by\cite{Kubiznak:2012wp,Abdusattar:2023hlj}
\be\begin{aligned}\label{criticalexponent}
    T\left(\frac{\partial S}{\partial T}\right)_V & \propto |\t-1|^{-\tilde{\a}} \equiv\tilde{\tau}^{-\tilde{\alpha}}\,,\\
   \frac{(\mR_A)_g-(\mR_A)_s}{\mR_c}&\propto|\tilde{\t}|^{\tilde{\beta}}\,,\\
    -\frac{1}{V}\left(\frac{\partial V}{\partial P}\right)_T &\propto |\tilde{\t}|^{\tilde{\g}}\,,\\
    p-1 &\propto(r_A-1)^{\tilde{\delta}}\equiv\omega^{\tilde{\delta}}
\end{aligned}\ee
Among them, $\tilde{\a}$ describes the behavior of specific heat capacity, and $\tilde{\g}$ describes the behavior of isothermal compressibility. $(\mR_A)_g$ and $(\mR_A)_s$ represent for the volume of two different phases. $\tilde{\delta}$ describes the behavior of the isothermal for $T=T_c$.
Since $S$ is just function of $V$, we can get $\tilde{\a}=0$. Then, expanding \eqref{dimensionlesseos} near the critical point($\tau=1$ and $\omega=0$) and ignoring higher-order term of $\omega^3$ and $\tau \omega$, we can obtain
\be\begin{aligned}\label{pexpand}
p^{(A)}=1+p_{10}^{(A)}\tilde{\t}+p_{11}^{(A)}\tilde{\t}\omega+p_{03}^{(A)}\omega^3\,,
\end{aligned}\ee
where $p_{ij}^{(A)}$ are constants shown in Appendix \ref{app}. The label $A$ represents model A. 
Following a similar procedure, using Maxwell's equal area law, we can get $\tilde{\beta}=\frac{1}{2}$. Differentiating \eqref{pexpand} gives $\tilde{\gamma}=0$. And taking $\tilde{\t}=0$ in \eqref{pexpand} gives $\tilde{\delta}=3$ directly. These values exactly match those of the van der Waals fluid and the mean-field theory of second-order phase transitions. Also, it is easy to check that the scaling law is satisfied in this model.

\subsection{Model B}\label{sec4-2}


The second model is a dynamic dark energy model based on GP\cite{Savaliya:2025cct}.
In this model, a Proca field is non-minimally coupled to spacetime curvature through dimensionless marginal coupling terms. 
It naturally yields an effective dark energy equation of state that depends on the Hubble parameter, reproducing the deviation from $\Lambda$CDM suggested by current observations such as DESI, while asymptotically approaching the standard de Sitter universe at late times. Meanwhile, the model is stable throughout the entire cosmic evolution\cite{Savaliya:2025cct}.

We will give up the ansatz that requires the time component Proca field $\phi^p$ proportional to $H^{-1}$. Following ref.\cite{Savaliya:2025cct}, we use the ansatz 
\be\begin{aligned}
    G_2=\mathcal{F}-\tilde{m}^2 X\,,\,\,\,G_3=\tilde{\gamma}_3 X\,,\,\,\, G_4=\frac{M_{pl}^2}{2} -\gamma_4 X\,,\,\,\, G_5=\gamma_5\,.
\end{aligned}\ee
Here, $\gamma_{3,4,5}$ are all dimensionless constants, and $m$ is the Proca mass parameter with dimension $[L^{-1}]$, which is the same as the Hubble parameter.
Then, solving the Proca equation of motion \eqref{eomA}, we can obtain
\be\begin{aligned}\label{model2phi}
    \phi=\frac{\tilde{m}^2+6\gamma_4 H^2}{3\tilde{\gamma}_3 H}\,.
\end{aligned}\ee
Although the $G_i$ functions are also in power-law form as discussed in \ref{sec4-1}, the two models are essentially different. For models with marginal couplings, $p$ can only take $-1/3,0,1$. The ansatz $\phi^p\propto H^{-1}$ contradicts the physical expectation that the energy density of the vector field should not overtake the background matter densities until the late-time universe for $p=-1/3,0$. And $\gamma_4$ and $\gamma_5$ will be dimensional for $p=1$, which is in contradiction to the marginal coupling\cite{Savaliya:2025cct}.

In order to connect with the observational restrictions given in ref.\cite{Savaliya:2025cct}, we would like to define the following dimensionless quantity and use them in this subsection:
\be\begin{aligned}
    \gamma_3=\frac{M_{pl}\tilde{\gamma}_3}{H_0}\,,\,\,\,m=\frac{\tilde{m}}{H_0}\,.
\end{aligned}\ee

Substituting \eqref{model2phi} into \eqref{friedmann}, we can acquire
\be\begin{aligned}
    \rho_m=&3 M_{pl}^2 H^2-\frac{\left(m^2 H_0^2-6 \gamma _4 H^2\right) \left(6 \gamma _4 H_0 H^2+H_0^3 m^2\right){}^2}{18 \gamma _3^2 H^2 M_{pl}^2},,\\
    p_m=&-M_{pl}^2 (2 \d H +3 H^2) \\
    &+\frac{H_0^2(6  \gamma _4 H^2+m^2 H_0^2)}{54 M_{pl}^2 \gamma _3^2 H^4 } \left[ -36  \gamma _4^2 \left(4 \d H H^4+3 H^6\right)+12 \gamma _4  H_0^2 \d H H^2 m^2+m^4 H_0^4 \left(3 H^2-2 \d H\right)\right]\,.
\end{aligned}\ee
This further gives $dE=-TdS+W dV$ with
\be\begin{aligned}
E=&4\pi M_{pl}^2\mR_A-\frac{2 \pi  \left(m^2 \mR_A^2 H_0^2-6  \gamma _4\right) \left(6 H_0 \gamma _4+m^2 \mR_A^2 H_0^3\right){}^2}{27 \gamma _3^2 \mR_A M_{pl}^2}\,,\\
S=&8M_{pl}^2 \pi^2 \mR_A^2+\frac{4 \pi ^2 m^6 \mR_A^6 H_0^8}{81 M_{pl}^2 \gamma _3^2}+\frac{16  \pi ^2 H_0^4 \gamma _4^2 m^2 \mR_A^2}{3 \gamma _3^2 M_{pl}^2}+\frac{128  \pi ^2 H_0^2 \gamma _4^3 \log (\mR_A/\mR_0)}{\gamma _3^2 M_{pl}^2}\,,\\
W=&-\frac{M_{pl}^2(\d \mR_A-3)}{  \mR_A^2}\\
&-\frac{6 \gamma _4+m^2 \mR_A^2 H_0^2}{54 M_{pl}^2 \gamma _3^2 \mR_A^4 }\left[36  \gamma _4^2 (2 \d\mR_A-3)-6  H_0^2 \gamma _4 \d\mR_A m^2 \mR_A^2+(\d\mR_A+3) m^4 \mR_A^4 H_0^4\right]\,,
\end{aligned}\ee
where $\mR_0$ is a integrate constant.
The temperature and volume is the same with the one given in \eqref{thermo1}. Similarly, we can choose the internal energy $U=-E$ and the pressure $P=W$ to get the thermodynamic first law of the apparent horizon in this model.
Then, it is easy to obtain the equation of state
\be\begin{aligned}\label{PTR2}
    P(T,\mR_A)=&\frac{M_{pl}^2(4 \pi  \mR_A T+1)}{  \mR_A^2}+\frac{m^6 H_0^8 \mR_A^2 (4 \pi  \mR_A T-5)}{54 M_{pl}^2 \gamma _3^2}\\
    &-\frac{\gamma _4 m^4  H_0^6}{3 M_{pl}^2 \gamma _3^2}+\frac{2 H_0^4 \gamma _4^2 m^2 (4 \pi  \mR_A T+1)}{3 M_{pl}^2 \gamma _3^2 \mR_A^2}+\frac{4 \gamma _4^3 H_0^2(8 \pi  \mR_A T-1)}{M_{pl}^2\gamma _3^2 \mR_A^4 }\,.
\end{aligned}\ee
It is hard to get an analytical result of the critical point. However, after giving the coupling constants $m$ and $\gamma_{3,4}$, we can calculate the solution of $\left(\partial P/\partial \mR_A\right)_T=(\partial^2 P/\partial \mR_A^2)_T=0$ numerically. If there are points where $T>0$ and $\mR_A>0$, it indicates that a phase transition exists. Conversely, if there is no positive solution to $T$ and $\mR_A$, then the phase transition does not occur. 

According to ref.\cite{Savaliya:2025cct}, the range of the value of the parameter constrained by observation is
\be\begin{aligned}
    -4.05\leq \log_{10} \gamma_3\leq -0.38\,,\,\,\,\gamma_4<2.98\times 10^{-11}\,,\,\,\,0.42\leq m\leq1.18\,.
\end{aligned}\ee
We employed the Monte Carlo method and randomly selected $10^5$ points within the parameter range to calculate the critical point. We did not find any parameters that could lead to an $P-V$ phase transition. Therefore, we would like to conclude that there will be no $P-V$ transition in model B.




\section{Conclusion}\label{sec5}

This study has explored the thermodynamics of the flat FRW universe in the context of GP theory. By considering two viable dark energy models constructed from the GP action, we have derived their respective thermodynamic descriptions on the apparent horizon, including the first law and the equation of state. We also studied the critical phenomena of the two models.

For Model A, which features power-law couplings and a specific relation between the Proca field and the Hubble parameter $\phi^p\propto H^{-1}$, we successfully derived an analytical equation of state in terms of the apparent horizon temperature and radius. Our analysis revealed a clear $P-V$ criticality, with a well-defined critical point whose existence depends on the sign of a model specific constant $C_1$. 
The resulting $P–V$ diagram displays the features of a van der Waals–type transition: isotherms below the critical temperature develop an S-shaped segment, signaling thermodynamic instability and the coexistence of two distinct phases.
At the critical isotherm, the curve exhibits a smooth inflection precisely at the dimensionless coordinates $(p,r_A)=(1,1)$, confirming second-order criticality.
Furthermore, we calculated the four critical exponents $(\tilde{\a},\tilde{\b},\tilde{\gamma},\tilde{\delta})$ and found them to be (0,1/2,0,3) , which are identical to those of a standard van der Waals fluid and satisfy the expected scaling laws. This suggests a universal character for the phase transition in this class of dark energy models.

By contrast, Model B, which relies on marginal curvature couplings and is tightly constrained by recent data from DESI, Planck, SNIa, and BAO, does not exhibit such critical behavior within its viable parameter region. Extensive numerical scans over the allowed ranges yield that equations of state lack physical critical points, i.e., no solutions with both positive temperature and horizon radius satisfying the necessary extremum conditions. 
This distinction illustrates how horizon thermodynamics can serve as a discriminating criterion beyond purely geometric or observational fits—offering insight into the internal consistency and physical richness of competing theoretical frameworks.

Our results demonstrate that the thermodynamic framework provides a powerful and insightful tool for analyzing the macroscopic behavior of the universe in vector-tensor theories like GP gravity, revealing deep connections between gravitational dynamics and thermodynamic phase structure.

\appendix
\section{Expressions of some parameter in the main text}\label{app}

The $C_i$ constant in \ref{sec4-1} is given by
\be\begin{aligned}
&C_1=   -3 \beta _2 H_0^{2+2s} \left(p \left(12 \beta _4 p_2-4 \beta _5 p_2-1\right)+p_2 \left(6 \beta _4 \left(2 p_2-1\right)-4 \beta _5 p_2-1\right)\right)\,,\\
&C_2= \beta _2  H_0^{2+2s} \left(4 \left(3 \beta _4-\beta _5\right) p_2^2+p_2 \left(6 \beta _4 (2 p-1)-4 \beta _5 p-1\right)-p\right)\,,
\end{aligned}\ee
Noting that $C_1$ can be written as 
\be\begin{aligned}
    C_1=-3\beta_2 H_0^{2+2s} \frac{\beta}{p_2}\,,
\end{aligned}\ee
where $\beta=p_2(\left(p \left(12 \beta _4 p_2-4 \beta _5 p_2-1\right)+p_2 \left(6 \beta _4 \left(2 p_2-1\right)-4 \beta _5 p_2-1\right)\right))$ is exactly the parameter describing the dark energy in eq.(5.8) of Ref.\cite{DeFelice:2016yws}.

The precise expression of the coefficient $\alpha_i$ in \eqref{criticalpoint} is given by
\be\begin{aligned}
    &\a_1=\frac{17^{3/4}}{2\cdot 3^{5/6} 5^{5/12} \sqrt[4]{7} \sqrt[12]{3154812 \sqrt{119}+34414919} \pi }\,,\\ 
    &\a_2=\frac{3^{5/6} 5^{5/12} \sqrt[12]{3154812 \sqrt{119}+34414919}}{\sqrt[4]{119}}\,,\\
    &\a_3=\frac{17^{5/6} \sqrt[6]{8565176 \sqrt{119}-81678387}}{100 \sqrt{5} 21^{2/3} \pi }\,.
\end{aligned}\ee

Then explicit expression of $p^{(A)}_i$ in \eqref{pexpand} is given by
\be\begin{aligned}
    &p^{(A)}_{10}=-\frac{\alpha _1 \left(\alpha _2^{12/5}-45\right)}{90 \alpha _2 \alpha _3}=\frac{1}{187} \left(68-8 \sqrt{119}\right)\,,\\
    &p^{(A)}_{01}=0\,,\\
    &p^{(A)}_{11}=-\frac{\alpha _1 \left(7 \alpha _2^{12/5}+225\right)}{450 \alpha _2 \alpha _3}=\frac{4}{935} \left(3 \sqrt{119}-119\right)\,,\\
    &p^{(A)}_{02}=0\,,\\
    &p^{(A)}_{03}=\frac{34 \alpha _2^{12/5}+\pi  \alpha _1 \left(7 \alpha _2^{12/5}-5625\right) \alpha _2-5625}{11250 \pi  \alpha _2^2 \alpha _3}=\frac{1}{825} (-14) \left(\sqrt{119}-3\right)\,.
\end{aligned}\ee

\acknowledgments
AS is supported by the NSFC (Grant No. 12250410250) and a Provincial Grant (Grant No. 2023QN10X389). AS would like to thank Jie Jiang for useful discussion.



\bibliographystyle{JHEP}
\bibliography{biblio.bib}




\end{document}